\documentstyle[aps,epsf]{revtex}

\begin{document}

\draft

\title{Relativistic Structure of the Nucleon Self-Energy\\
in Asymmetric Nuclei}
 
\author{S.\ Ulrych and H.\ M\"{u}ther} 
\address{Institut f\"ur Theoretische Physik, Universit\"at T\"ubingen,
         D-72076 T\"ubingen, Germany}

\maketitle

\begin{abstract}
The Dirac structure of the nucleon self-energy in asymmetric nuclear matter
cannot reliably be deduced from the momentum dependence of the single-particle
energies. It is demonstrated that such attempts yield an isospin
dependence with even a wrong sign. Relativistic studies of finite nuclei have
been based on such studies of asymmetric nuclear matter. The effects of these
isospin components on the results for finite nuclei are investigated. 
\end{abstract}


\section{Introduction}
During the last years a substantial progress has been made in the 
microscopic description of the bulk properties of nuclear matter by the
inclusion of relativistic features in the so-called
Dirac-Brueckner-Hartree-Fock (DBHF) approximation. One finds that the
self-energy of the nucleon in nuclear matter contains a large attractive
component $\Sigma_s$ which is of the order of -300 MeV and transforms like 
a scalar under a Lorentz transformation. This very attractive contribution is
compensated to a large extent by a repulsive, time-like Lorentz-vector 
component $\Sigma_0$. This partial cancellation between $\Sigma_s$ and
$\Sigma_0$  leads to single-particle energies and binding energies for the
nucleons, which are of the order of -40 MeV, small compared to the nucleon 
rest-mass. Based on this small binding energy it has been argued for a long time
that relativistic effects should be small in nuclear physics.

However, considering a self-energy with a relativistic structure as just
outlined in a Dirac equation for a nucleon in a medium of nuclear matter, one
finds that Dirac spinors derived from this equation exhibit a substantial
enhancement of the small component as compared to the Dirac spinor of a free
nucleon with the same momentum. This Dirac spinor essentially corresponds to one
for a nucleon with an effective mass $m^*$ which is the sum of the bare mass $m$
plus the scalar part of the self-energy $\Sigma_s$. Assuming a value for
$\Sigma_s$ of -300 MeV, which is quite typical for nuclear matter at saturation
density, it is obvious that this reduction of the effective mass can result in
non-negligible effects. The matrix elements of the nucleon-nucleon (NN)
interaction for two nucleons in nuclear matter should be evaluated employing 
these Dirac spinors modified by the nuclear medium, which means that the NN
interaction is density-dependent. It is this density-dependence which leads to
the saturation of nuclear matter in simple relativistic mean-field calculations
within the Walecka model\cite{serot}.

Replacing the phenomenological approach for the NN interaction used in the
Walecka model by a realistic meson-exchange potential, one has to perform a
nuclear structure calculation which goes beyond the mean-field or Hartree-Fock
approximation and account for the effects of NN correlations. This can be done
in a relativistic extension of the Brueckner-Hartree-Fock approximation. For
that purpose one has to evaluate the $G$-matrix by solving the Bethe-Goldstone
equation. For relativistic models of the NN interaction like the
One-Boson-Exchange (OBE) potentials this Bethe-Goldstone equation corresponds to
a three-dimensional reduction of the Bethe-Salpeter equation, which accounts for
Pauli- and dispersive corrections due to the surrounding nuclear
medium\cite{rupr}. 

The $G$-matrix derived from this Bethe-Goldstone equation can be analysed and
decomposed into five Lorentz invariant amplitudes. From these amplitudes one can
calculate the nucleon self-energy within the Brueckner-Hartree-Fock 
approximation and determine its relativistic structure, which means its 
decomposition into a scalar term $\Sigma_s$, a time-like vector term 
$\Sigma_0$ and a space-like vector term $\Sigma_v$. This self-energy is included
in a Dirac equation to determine the Dirac spinors for the nucleons. A
self-consistent solution of these DBHF equations requires that the resulting
spinors are used to evaluate the matrix elements of the OBE interaction and
determine the $G$-matrix. 

Such self-consistent DBHF calculations for nuclear matter have been performed by
various groups employing different models for the NN interaction 
\cite{rupr,anast,brock,malf1,weigel}. All these investigations show that the
relativistic effects modify the saturation properties of
nuclear matter derived from realistic NN interaction within the BHF scheme. The
density-dependence of the nucleon Dirac spinors yields some repulsion which
increases significantly with the nuclear density. Due to this mechanism the
Dirac effects of the DBHF approach supply a fine-tuning in the calculated energy
of nuclear matter as  a function of density, which moves the prediction for the
saturation point off the well-known Coester band\cite{coester}, which is 
obtained in non-relativistic many-body calculations of nuclear matter, towards
the empirical result. Brockmann and Machleidt actually succeeded in finding a
version of the Bonn potential, which fits the NN scattering data and reproduces
the empirical saturation point of symmetric nuclear matter using the DBHF
approach\cite{brock}.

Various groups analyzed the relativistic structure of the $G$-matrix and derived
the three components of the self-energy, $\Sigma_s$, $\Sigma_0$ and $\Sigma_v$,
in symmetric nuclear matter. They find the dependence of these three components 
on the momentum of the nucleon is rather weak and that the effects of the
space-like vector component $\Sigma_v$ is rather small as compared to the other
two\cite{malf1,weigel,elsen,boer1}. Therefore it seems to be justified that one
takes advantage of this feature and derives the decomposition of the self-energy
into $\Sigma_s$ and $\Sigma_0$ from the momentum-dependence of the 
single-particle energy $\epsilon (k)$ (see also below). In this way one can
avoid the analysis of the relativistic structure of $G$, which simplifies the
self-consistent procedure of DBHF significantly.

After the DBHF scheme had been applied with great success to symmetric nuclear
matter, it was an obvious extension to use the same scheme for pure neutron
matter and assymetric nuclear matter with various fractions of proton to neutron
density as well\cite{prak,engv1,engv2}. In section 2, below, 
we would like to demonstrate
that it is very dangerous to apply the simplified version of the DBHF
self-consistency, i.e.~deducing the Dirac components from the single-particle
energies, to asymmetric nuclear systems. We will see that this scheme, which has
been used e.g.~by \cite{prak,engv1,engv2,kuo} even tends
to predict the ``wrong sign'' for the isovector dependence of $\Sigma_s$ and
$\Sigma_0$.

Attempts have also been made to apply the DBHF approach to finite nuclei as
well\cite{mut1,fritz,toki1,toki2,boer2}. One possibility in this direction
is the so-called Relativistic Mean Field approach with Density dependent
coupling constants (RMFD)\cite{toki1,toki2}. The first step of this approach
considers the Relativistic Mean Field or Dirac-Hartree approach for nuclear
matter at a given density, adjusting an effective coupling constant for the 
scalar meson $\sigma$ and the vector meson $\omega$ to reproduce the results of
microscopic DBHF calculations at this density. In a second step these 
density-dependent coupling constants are employed in a Dirac-Hartree calculation
to evaluate the ground-state properties of finite nuclei in a kind of
local-density approximation. 

Recently this RMFD approach has been extended to asymmetric nuclear systems.
Using the DBHF results of Engvik et al.\cite{engv2}, Shen et al.\cite{toki2}
determined density-dependent coupling constants for the isoscalar mesons
$\sigma$ and $\omega$ and the isovector mesons $\delta$ and $\rho$ to fit the 
relativistic components of the self-energy for protons and neutrons at various
densities and asymmetries. Since, however, this analysis is based on DBHF
calculations which exhibits the ``wrong sign'' in the isospin structure, one 
obtains coupling constants for the isovector mesons in particular, which are not
appropriate. In section 3 we would like to correct this sign and show the 
effect of a proper treatment of the isovector mesons on the structure of 
finite nuclei using the RMFD approach.

\section{Isovector mesons and asymmetric nuclear matter}
Our investigation of the role of isovector mesons in the effective NN
interaction in a nuclear medium is based on the analysis of Boersma and
Malfliet\cite{boer1}. They parametrize the relativistic structure of the
G-matrix in terms of five Lorentz invariants
\begin{equation}
G = \sum_{\alpha=1}^5 T^{\alpha} F^{\alpha}_{(1)} F^{\alpha}_{(2)} \, ,
\label{eq:para1}
\end{equation} 
with
\begin{equation} 
 F^{\alpha}_{(i)} = \left\lbrace 1,\,  \sigma^{\mu\nu}, \,
\gamma_5\gamma^\mu,\, \gamma^\mu,\, \gamma_5 \frac{\not\! q}{2m^*}\right\rbrace
\, . \label{eq:para2}
\label{equation}
\end{equation}
The amplitudes $T^{\alpha}$ are parametrized in terms of Yukawa functions,
depending on the Mandelstam variable $t$ as
\begin{equation}
T^\alpha = \sum_{n=1}^4 \frac{g_{\alpha n}^2}{\mu_n^2 - t}
\label{eq:para3}
\end{equation}
with effective meson masses $\mu_n$ and coupling constants $g_{\alpha n}$.
At a given density of nuclear matter, these coupling constants are adjusted in
such a way that the antisymmetrized matrix elements of the parametrization
(\ref{eq:para1}) - (\ref{eq:para3}) reproduce the corresponding matrixelements
of the $G$-matrix, calculated in the rest frame of nuclear matter\cite{dejong}.
More details of this parametrization can be found in \cite{boer1,boer3}.
All the results displayed in this manuscript are obtained using the parameter
set III of \cite{boer1,boer3}. Results derived from parameter sets I and II are
rather similar.

Using this parametrization one can easily determine the self-energy $\Sigma^i 
(k)$ for protons and neutrons, labeled by index $i$, with momentum $k$
\begin{equation}
\Sigma^i (k) = 1 \tilde{\Sigma^i_s} (k) + \gamma \cdot {\bf k} \Sigma_v^i (k) - 
\gamma^0 \tilde{\Sigma_0^i} (k)\, , \label{eq:self}
\end{equation} 
by solving the Dirac-Hartree-Fock equation for the NN interaction defined by the
parametrization of  (\ref{eq:para1}) - (\ref{eq:para3}) in a selfconsistent way.
Inserting this self-energy into the Dirac equation for a nucleon in the nuclear
medium, we obtain
\begin{equation}
\left[ \left( 1+\Sigma_v^i (k)\right) \gamma \cdot {\bf k} + \left( M +
\tilde{\Sigma^i_s} (k) \right) - \tilde{\Sigma_0^i} (k)  \gamma^0 \right] 
u_i (k) = \epsilon_i (k) \gamma^0 u_i (k)
\label{eq:dirac1}
\end{equation}
Now it is convenient to eliminate $\Sigma_v$ and rewrite this Dirac equation 
into a form which only contains a scalar and a time-like vector component
\begin{equation}
\left[  \gamma \cdot {\bf k} + \left( M +
{\Sigma^i_s} (k) \right) - {\Sigma_0^i} (k)  \gamma^0 \right] 
u_i (k) = \epsilon_i (k) \gamma^0 u_i (k)
\label{eq:dirac2}
\end{equation}
where
\begin{eqnarray}
\Sigma_s^i & = & \frac{\tilde{\Sigma_s^i} - M \Sigma_v^i}{1+\Sigma_v^i}
\nonumber \\
\Sigma_0^i & = & \frac{\tilde{\Sigma_0^i} - \epsilon_i \Sigma_v^i}{1+\Sigma_v^i}
\label{eq:efself}
\end{eqnarray}

We will assume that the parametrization of $G$, which has been determined by
Boersma and Malfliet for symmetric nuclear matter at various densities $\rho$
may also be used for asymmetric nuclear matter at the same density. 
With this assumption we ignore the fact that the $G$-matrix will not only 
depend on the density of nuclear matter but also on the asymmetry parameter
\begin{equation}
\alpha = \frac{\rho^p}{\rho^n + \rho^p} \label{eq:defalpha}
\end{equation} 
where $\rho^p$ and $\rho^n$ denote the density of protons and neutrons,
respectively. This implies that symmetric nuclear matter corresponds to $\alpha
= 0.5$ and pure neutron matter to $\alpha = 0$. The assumption that the
$G$-matrix does not depend on this asymmetry parameter $\alpha$ may not be
sufficient for a very sophisticated study of asymmetric systems. It should be
sufficient, however, for the more general remarks, which we wish to make in the
present study.

\begin{figure}
\epsfysize=7.0cm
\begin{center}
\makebox[16.6cm][c]{\epsfbox{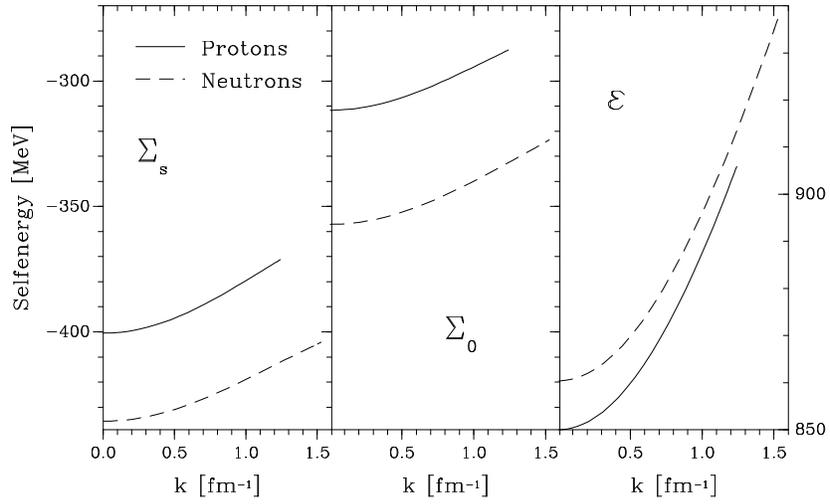}}
\end{center}
\caption{Scalar ($\Sigma_s$, left part) and vector component ($\Sigma_0$, middle
part) of the self-energy and the single-particle energy ($\epsilon$, see
(\protect\ref{eq:epsi} ), right part of the figure) for protons (solid line) and
neutrons (dashed line) as a function of momentum $k$. Results are displayed for
asymmetric nuclear matter with a baryon density $\rho$ = 0.185 nucleon
fm$^{-3}$ and an asymmetry parameter $\alpha$ = 0.35.}
\label{fig:one}
\end{figure}

As a typical example we will now first consider the case of asymmetric nuclear 
matter which is defined by a baryon density $\rho$ = 0.185 nucleon fm$^{-3}$,
which corresponds roughly to the saturation density of nuclear matter and an
asymmetry parameter $\alpha$ = 0.35. Results for the self-energy of protons and
neutrons are displayed in Fig.~\ref{fig:one} as a function of the momentum $k$.
One observes a momentum dependence of the scalar and vector components,
$\Sigma_s^i$ and $\Sigma_0^i$ defined in (\ref{eq:efself}), which is rather weak 
as compared to the total value of these components. The variation of these
quantities as a function of momentum is around 20 MeV for momenta below the
Fermi momentum, which corresponds to about 5 percent of the total value.
Therefore one may be tempted to ignore this momentum-dependence for a moment and
interpret the mean value for these components
\begin{equation}
U_\beta^i = \frac {\int_0^{k_{Fi}} k^2 \Sigma_\beta^i (k)
\,dk}{\frac{k_{Fi}^3}{3}}\, ,
\label{eq:mean}
\end{equation}
in terms of a mean field or Hartree-Fock model. The index $\beta$ in this 
equation represents the scalar ($\beta =s$) or vector component ($\beta = 0$)
and $k_{Fi}$ stands for the Fermi momentum of protons ($i=p$) and neutrons
($i=n$), respectively. Assuming a meson exchange model for the NN interaction,
which considers the exchange of a scalar meson ($\sigma$) and a vector meson
($\omega$) plus the exchange of the corresponding isovector mesons $\delta$ and
$\rho$, the components of the self-energy for protons and neutrons are
easily evaluated within the Hartree approximation as
\begin{eqnarray}
U_s^p & = & -\frac{g_\sigma^2}{m_\sigma^2} \left( \rho_s^p + \rho_s^n \right) 
-\frac{g_\delta^2}{m_\delta^2} \left( \rho_s^p - \rho_s^n \right)\nonumber\\
U_s^n & = & -\frac{g_\sigma^2}{m_\sigma^2} \left( \rho_s^p + \rho_s^n \right) 
+\frac{g_\delta^2}{m_\delta^2} \left( \rho_s^p - \rho_s^n \right)\nonumber\\
U_0^p & = & -\frac{g_\omega^2}{m_\omega^2} \left( \rho^p + \rho^n \right)
- \frac{g_\rho^2}{m_\rho^2}\left( \rho^p - \rho^n \right)\nonumber\\
U_0^n & = & -\frac{g_\omega^2}{m_\omega^2} \left( \rho^p + \rho^n \right)
+ \frac{g_\rho^2}{m_\rho^2}\left( \rho^p - \rho^n \right) \label{eq:hartre}
\end{eqnarray}
The parameter $g_\mu$ and $m_\mu$ in these equations refer to the meson-nucleon
coupling constants and the masses of the varios mesons, the proton and neutron
densities are represented by $\rho^p$ and $\rho^n$ and the scalar densities are
defined by
\begin{equation}
\rho_s^i = \frac{8\pi}{(2\pi )^3}\int_0^{k_{Fi}} k^2 \frac{M^*_i}{E^*_i} dk
\label{eq:rhos}
\end{equation}
with the effective mass
\begin{equation}
M^*_i = M + U_s^i \label{eq:mstar}
\end{equation}
and
$$
E^*_i = \sqrt{M^*_i + k^2}\, .
$$
Inserting these Hartree-results for the self-energy into the Dirac
eq.(\ref{eq:dirac2}) one obtains the single-particle energy in the Hartree
approximation
\begin{equation}
\epsilon_i^H (k) = \sqrt{M^*_i + k^2} - U_0^i\, . \label{eq:epshart}
\end{equation} 
For nuclear matter with $\rho^n > \rho^p$ the Hartree approximation of 
(\ref{eq:hartre})
predicts more negative values for the self-energy components for neutrons,
$U_0^n$ and $U_s^n$, as compared to the corresponding results for the protons.
Looking again at the momentum-dependent results for these self-energy components
displayed in Fig.~\ref{fig:one} we observe that this feature is reproduced 
by the momentum-dependent self-energy components derived from the Groningen
parametrization of DBHF. From this figure we can furthermore observe that the
isospin dependence is larger for the vector component of the self-energy 
$\Sigma_0$ than for the scalar component. Translated into the Hartree analysis 
of eq.(\ref{eq:hartre}) this means that the effective $\rho$ exchange
contribution to the $G$ matrix seems to be stronger than the effects of the
isovector scalar meson $\delta$. This is definetely true for a realistic
meson-exchange potential like the Bonn potential\cite{rupr}, and obviously
remains valid for the $G$ matrix. The difference in the isospin dependence of
$\Sigma_s$ and $\Sigma_0$ is also responsible for the single-particle energies
\begin{equation}
\epsilon_i (k) = \sqrt{\left( M+\Sigma_s^i (k)\right)^2 + k^2} - \Sigma_0^i (k)
\label{eq:epsi}
\end{equation}
and leads to the result that the single-particle energies for the protons are
more attractive than those for the neutrons (see part on the right-hand side of
Fig.~\ref{fig:one}).  This reflects the fact that the proton - neutron
interaction is more attractive than the neutron - neutron interaction.

\begin{figure}
\epsfysize=7.0cm
\begin{center}
\makebox[16.6cm][c]{\epsfbox{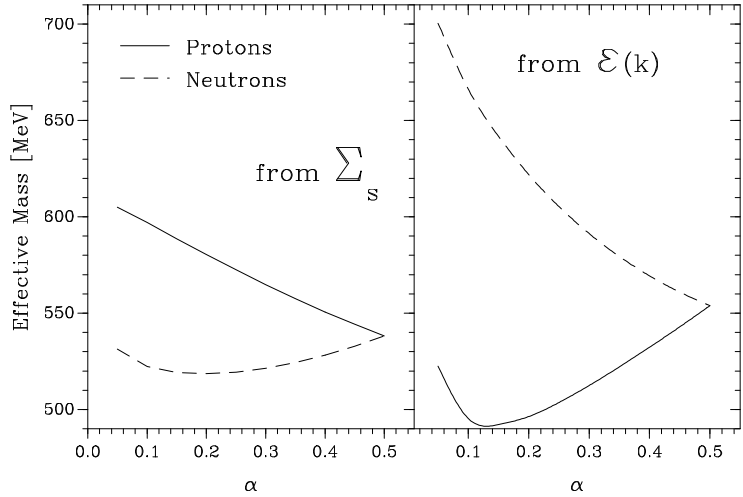}}
\end{center}
\caption{Effective masses of nuclear matter ($\rho$ = 0.185 nucleon fm$^{-3}$)
as a function of the asymmetry parametr $\alpha$. The effective masses displayed
in the left part of the figure have been derived from the momentum averaged
self-energy $\Sigma_s$ of eq.(\protect \ref{eq:mean}), while those in the right
part of the figure are deduced from the single-particle energies $\epsilon (k)$
according eq.(\protect \ref{eq:epshart} ).}
\label{fig:two}
\end{figure}

One of the main features of the dependence of the self-energy for protons and
neutrons in asymmetric nuclear matter is displayed in the left part of
Fig.~\ref{fig:two}. There we show the dependence of the average effective mass
$M_i^*$ calculated according eq.(\ref{eq:mstar}) using the momentum averaged
$U_s^i$ defined in (\ref{eq:mean}) as a function of the asymmetry parameter 
$\alpha$. The effective mass is of course identical for protons and neutrons in
the case of symmetric nuclear matter ($\alpha$ =0.5). The effective mass for the
protons increases with decreasing $\alpha$ i.e. with an increasing fraction of
neutrons. The effective mass for the neutrons is smaller than the corresponding
$M_p^*$ for all values $0 \le \alpha < 0.5$. 

This is just opposite to the behaviour observed by Engvik et
al.\cite{engv2,engv3} (see also Fig.2 of \cite{toki2}) as well as Lee et 
al.\cite{kuo}. These investigations report effective masses for the proton to be
smaller than those for the neutron. One could argue that this different
behaviour of $\Sigma_s^i$ or $M^*_i$ is due to the different interactions: the
results of \cite{engv2} have been derived from the Bonn potential A, whereas we
consider parametrization of the Groningen potential. However, we would like to
demonstrate that the isospin dependence of the effective mass $M^*_i$ reported
in \cite{engv2} is wrong and that this mistake is caused by the fact that Engvik
et al.~deduce the effective mass from the momentum dependence of the
single-particle energy, a method which has been used by other groups before
\cite{brock,prak,kuo}.

In order to demonstrate this, we consider the momentum dependent self-energies
$\Sigma_{s,0}^i (k)$ deduced from the Groningen potential and calculate the
single-particle energy $\epsilon_i (k)$ calculated according to (\ref{eq:epsi}).
In a next step we analyze these functions $\epsilon_i (k)$ and deduce from these
functions effective masses $M_i^*$ and vector potentials $U_0^i$. This can be
done by taking $\epsilon_i$ at two values of $k=k_1,k_2$ (our choice is $k_1 =
0.7 k_F$, $k_2 = k_F$) and by adjusting the
parameters $M_i^*$ and $U_0^i$ in such a way that (\ref{eq:epshart}) reproduces
the values $\epsilon_i (k_1)$ and $\epsilon (k_2)$. In this way we simulate the
procedure of Engvik et al. \cite{engv2} and Lee et al. \cite{kuo} for 
the Groningen potential. 

The results of this procedure is displayed in the right part of
Fig.~\ref{fig:two}. We find that the effective masses deduced from the momentum
dependence of the single-particle energies show the same behaviour as reported
by Engvik et al. \cite{engv2} and Lee et al. \cite{kuo}. This demonstrates very
clearly that the isospin dependence dependence of the effective mass is very
sensitive to the method used to determine it. Although the momentum dependence
of the components of the self-energy $\Sigma_{0,s}^i$ is rather weak on the
scale of the absolute values of these quantities (see discussion of
Fig.~\ref{fig:one} above), it is obviously too strong to neglect it in analyzing
the momentum-dependence of $\epsilon_i (k)$. Therefore such an analysis leads to
wrong results, in particular when we are interested in the isospin dependence. 

\begin{figure}
\epsfysize=7.0cm
\begin{center}
\makebox[16.6cm][c]{\epsfbox{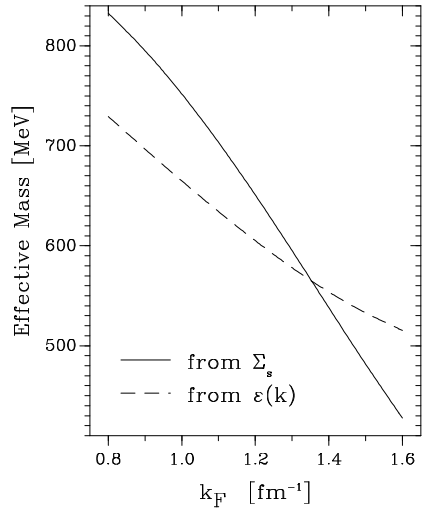}}
\end{center}
\caption{Effective masses of symmetric nuclear matter as a function of the Fermi
momentum $k_F$. The results represented by the solid line have been derived from
the momentum averaged self-energy $\Sigma_s$ of eq.(\protect \ref{eq:mean}).
Those shown by the dashed line are deduced from the single-particle energies
$\epsilon (k)$ according eq.(\protect \ref{eq:epshart} ).}
\label{fig:twoa}
\end{figure}

This observation led us to compare the two schemes of analyzing the
Dirac structure of the nucleon also in the case of symmetric nuclear matter. 
Results for the effective mass as a function of the Fermi momentum $k_F$ of 
symmetric nuclear matter are displayed in Fig.~\ref{fig:twoa}. The results are
rather similar for densities around the empirical saturation density ($k_F
\approx 1.4$ fm$^{-1}$). This suggests that the evaluation of the saturation
properties of nuclear matter, as performed by Brockmann and Machleidt
\cite{brock} may not be influenced too much by improving the method to determine
the Dirac structure of the nucleon spinors. For other densities, however, 
the effective mass deduced from the momentum dependence of the single-particle 
energy (dashed line) deviates significantly from the effective mass derived 
directly from $\Sigma_s$. In particular at smaller densities, these are the 
densities which are relevant for the description of finite nuclei with medium
mass, the effective mass derived from the single-particle energies is
considerably smaller than the corresponding value deduced from $\Sigma_s$.  
  
\section{Relativistic mean field calculations for finite nuclei}  
Relativistic calculations for finite nuclei have often been based on DBHF 
studies of nuclear matter using a kind of local density approximation
\cite{mut1,fritz,toki1,toki2,boer2}. One possibility is to consider asymmetric
nuclear matter at a given density $\rho$ and asymmetry $\alpha$ and determine
the scalar and vector components $U_s$ and $U_0$ of the nucleon self-energy for
protons and neutrons. Using the set of equations (\ref{eq:hartre}) one can
determine effective coupling constants $g_\sigma^2$, $g_\omega^2$, $g_\delta^2$,
and $g_\rho^2$. This means that a Dirac-Hartree calculation of nuclear matter  
allowing for the exchange of $\sigma$, $\omega$,
$\delta$, and $\rho$ mesons with these coupling constants would reproduce the
results of the DBHF calculation at this density and asymmetry. These effective
coupling constants depend on density and asymmetry. 

\begin{figure}
\epsfysize=7.0cm
\begin{center}
\makebox[16.6cm][c]{\epsfbox{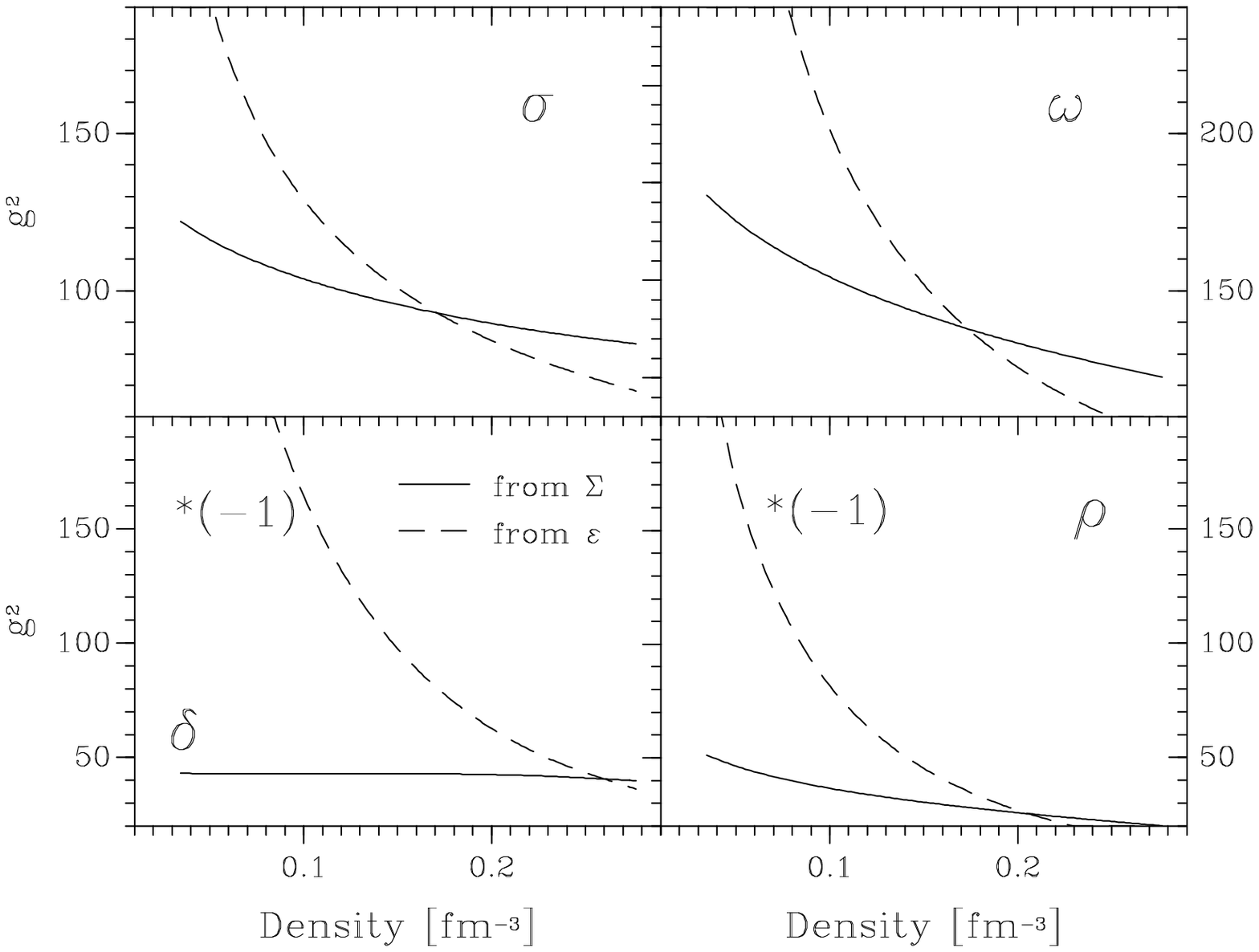}}
\end{center}
\caption{Effective coupling constants to be used in Dirac-Hartree calculations
as a function of density. These coupling constants 
are derived from (\protect\ref{eq:hartre}) assuming 
meson masses of 550 MeV, 783 MeV, 983 MeV, and 769 MeV for $\sigma$, $\omega$,
$\delta$, and $\rho$ mesons, respectively. The solid lines are derived from
the momentum averaged self-energies, while the dashed lines are obtained using
effective masses derived from the single-particle spectrum $\epsilon(k)$. 
Note that the
coupling constants $g_i^2$ are negative for the isovector mesons if they are
derived from $\epsilon (k)$.}
\label{fig:three}
\end{figure}

As an example we present in Fig.~\ref{fig:three} the density-dependence of these
coupling constants, derived from the mean values calculated according
(\ref{eq:mean}). These coupling constants are represented by the solid lines and
have been evaluated at an asymmetry defined by $\alpha$ =0.35. The dependence of
these coupling constants on $\alpha$ turns out to be weak. One can see that
these effective coupling constants exhibit a moderate density-dependence. The
coupling constants for the isovector mesons $\delta$ and $\rho$ are weaker than
those of the isoscalar mesons.

Instead of using the momentum averaged self-energies (\ref{eq:mean}) one may
also try to use the corresponding quantities $U_{s,0}^i$ derived from the
momentum-dependence of the single-particle energies $\epsilon (k)$. Using these
quantities as input to determine the effctive meson-coupling constants, we 
obtain the results represented by the dashed line. It must be mentioned that the
coupling constans $g^2$ for the isovector mesons turn out to be negative in
this analysis, which means that the coupling constants $g$ themselves would be
imaginary. This is a consequence of the ``wrong'' isospin dependence of these
self-energy components. The effective coupling constants derived from the
single-particle spectra also show a much stronger density dependence and the
absolute values for the isoscalar mesons are close to those of the isovector
mesons. We would like to mention again that both sets of effective coupling
constants (represented by the solid and dashed lines in Fig.~\ref{fig:three})
are determined for the same parameterization of the Groningen $G$-matrix. The
only difference is that the solid line is derived from the self-energies
directly, while the dashed lines are obtained from the attempt to derive these
self-energies from the single-particle spectrum.

Shen et al.\cite{toki2} determined effective coupling constants in the same way
as we just outlined using as input data the results of Engvik et
al.\cite{engv2}. This means that they start from components $U_{s,0}^i$ which
are derived from the single-particle spectrum $\epsilon (k)$ calculated for the
Bonn potential. It is remarkable that the coupling constants derived in
\cite{toki2} show features very similar to those represented by the dashed
line above. The coupling constants $g^2$ are negative for the isovector mesons
(note that the isovector components in eq.(6) of \cite{toki2} have a wrong sign)
and all effective coupling constants show a strong density dependence, very 
similar to the dashed lines in our Fig.~\ref{fig:three}. 

Considering a fixed asymmetry parameter $\alpha$ one may now consider the
various coupling constants, which for the asymmetry assumed just depend on the
total baryon density and solve the Dirac-Hartree or relativistic mean field
equations assuming a local density approximation for the meson-nucleon coupling
constant. Details of the procedure have been presented in \cite{fritz}. Note
that we do not include the effect of rearrangement terms \cite{toki2}. 
 
Results for such Dirac Hartree calculations with density dependent meson nucleon
coupling constants are listed in Tab.~\ref{tab:tab1}. As examples we consider
the ground-state properties of the nuclei $^{16}O$ and $^{22}O$. Three different
models are considered. (a): a pure $\sigma$ - $\omega$ model, the density
dependent coupling constants for these two mesons are derived from the results
of symmetric nuclear matter. (b): a $\sigma$, $\omega$, $\rho$ - $\delta$ model
in the limit of symmetric matter, i.e.~$\alpha$ = 0.5. The density dependent
coupling constants for these mesons are obtained by analyzing the proton and
neutron self-energies for asymmetric nuclear matter, extrapolated to the
symmetric case of $\alpha$ = 0.5. (c):  a $\sigma$, $\omega$, $\rho$ - $\delta$
model for asymmetric nuclear matter at $\alpha$ = 0.36, which corresponds the
proton fraction of $^{22}O$. For each of these models (a) - (c) the coupling
constants derived either from the relativistic structure of the self-energy
(results listed in the left part of the table) or from the momentum dependence
of the single-particle energies (results in the right part). 

One can see from these results that the two methods to determine the density
dependence of the effective meson coupling constants lead to quite different
results even for the isospin symmetric system $^{16}O$. The difference in the
calculated binding energy is as large as 0.6 MeV per nucleon.  The deviation of
the second scheme from the first one is even larger for the non-symmetric
nucleus $^{22}O$. Note that the differences between the two methods, around 0.9
MeV per nucleon, is larger than the effect of the isovector mesons: the
difference between the $\sigma-\omega$ model and the models including isovector
mesons is only around 0.3 MeV per nucleon.

\section{Conclusions}
The accuracy of the simple technique which tries to extract the Dirac structure
of the nucleon self-energy in nuclear matter from the momentum dependence of the
single-particle energy\cite{brock,prak,engv1,engv2,kuo} has been investigated 
using the relativistic parametrization of the $G$-matrix of Boersma et
al.~\cite{boer1}. It is demonstrated that the analysis based on the
single-particle spectrum fails if asymmetric nuclear systems are considered. The
isovector components of the self-energy derived in this way exhibit even the
wrong sign. The simplified analysis reproduces the structure of the self-energy
around the saturation density but overestimates the scalar part of the
self-energy at smaller densities. Therefore  the improvement of this scheme may
not have a very significant effect on the calculated saturation property of
symmetric nuclear matter \cite{brock}. A more apropriate analysis of the Dirac
structure of the nuclear self-energy is required to derive density-dependent
effective coupling constants to be used in Dirac-Hartree calculations of finite
nuclei.

This work has been supported by the ``Graduiertenkolleg Struktur und
Wechselwirkung von Hadronen und Kernen'' (DFG GRK 132/2).

\vfil\eject
\begin{table}
\begin{tabular}{|c c | c c | c c |}
&& \multicolumn{2}{c|}{from $\Sigma$}&\multicolumn{2}{c|}{from $\epsilon(k)$}\\
&& E/A [MeV] & $R_{Ch}$ [fm] & E/A [MeV] & $R_{Ch}$ [fm] \\
\hline
&&&&&\\
$^{16}O$& $\sigma ,\omega$ & -6.94 & 2.56 & -6.36 & 2.60 \\
& Exp. & -7.98 & 2.74 & & \\
&&&&&\\
\hline
&&&&&\\
$^{22}O$ & $\sigma ,\omega$ & -6.72 & 2.56 & -5.83 & 2.59 \\
& $\alpha = 0.5$ & -6.33 & 2.59 & -5.47 & 2.61 \\
& $\alpha = 0.36 $ & -6.40 & 2.58 & -5.35 & 2.63 \\
& Exp. & -7.36 & & & \\
&&&&&\\
\end{tabular}
\caption{Energy per nucleon (E/A) and radius of the charge distribution $R_{Ch}$
calculated in various versions of the Dirac Hartree approach with density
dependent coupling constants. The calculated energy has been corrected to
account for the effects of the spurious center of mass motion. Details of the 
various approaches are discussed in the text.}
\label{tab:tab1}
\end{table}

\vfil\eject

\begin{thebibliography}{99}
\bibitem{serot} B.D.\ Serot and J.D.\ Walecka, Adv.\ Nucl.\ Phys.\ {\bf 16}, 1
(1986)
\bibitem{rupr} R.\ Machleidt, Adv.\ Nucl.\ Phys.\ {\bf 19}, 189 (1989)
\bibitem{anast} M.R.\ Anastasio, L.S.\ Celenza, W.S.\ Pong, and C.M.\ Shakin,
Phys.\ Rep.\ {\bf 100}, 327 (1983)
\bibitem{brock} R.\ Brockmann and R.\ Machleidt, Phys.\ Lett.\ {\bf B149}, 283
(1984)
\bibitem{malf1} B.\ Ter Haar and R.\ Malfliet, Phys.\ Rep.\ {\bf 149}, 207
(1987)
\bibitem{weigel} H.\ Huber, F.\ Weber, and M.K.\ Weigel, Phys.\ Lett.\ {\bf
B317}, 485 (1993) 
\bibitem{coester} F.\ Coester, S.\ Cohen, B.D.\ Day and C.M.\ Vincent, Phys.\
Rev.\ {\bf C1}, 769 (1970)
\bibitem{elsen} H.\ Elsenhans, H.\ M\"uther, and R.\ Machleidt, Nucl.\ Phys.\
{\bf A515}, 715 (1990)
\bibitem{boer1} H.F.\ Boersma and R.\ Malfliet, Phys.\ Rev.\ {\bf C49}, 233
(1994), Erratum {\bf C50}, 1253 (1994)
\bibitem{prak} H.\ M\"uther, M.\ Prakash, and T.L.\ Ainsworth, Phys.\ Lett.\ {\bf
B199}, 469 (1987)
\bibitem{engv1} L.\ Engvik, M.\ Hjorth-Jensen, E.\ Osnes, G.\ Bao and
E.\ \O stgaard, Phys.\ Rev.\ Lett.\ {\bf 73}, 2650 (1994)
\bibitem{engv2} L.\ Engvik, E.\ Osnes, M.\ Hjorth-Jensen, G.\ Bao and
E.\ \O stgaard, Ap.\ J.\ {\bf 469}, 794 (1996)
\bibitem{kuo} C.-H.\ Lee, T.T.S.\ Kuo, G.Q.\ Li, and G.E.\ Brown, preprint
nucl-th/9703034
\bibitem{mut1} H.\ M\"uther, R.\ Machleidt, and R.\ Brockmann, Phys.\ Rev.\ {\bf
C42}, 1981 (1990)
\bibitem{fritz} R.\ Fritz and H.\ M\"uther, Phys.\ Rev.\ {\bf C49}, 633 (1994)
\bibitem{toki1} R.\ Brockmann and H.\ Toki, Phys.\ Rev.\ Lett.\ {\bf 68}, 3408
(1992)
\bibitem{toki2} H.\ Shen, Y.\ Sugahara, and H.\ Toki, Phys.\ Rev.\ {\bf C55},
1211 (1997) 
\bibitem{boer2} H.F.\ Boersma and R.\ Malfliet, Phys.\ Rev.\ {\bf C49}, 1495
(1994)
\bibitem{dejong} F.\ de Jong, Thesis, Groningen 1992, Phys.\ Rev. {\bf C44}, 998
(1991)
\bibitem{boer3} H.F.\ Boersma, Thesis, Groningen 1992
\bibitem{engv3} L.\ Engvik, M.\ Hjorth-Jensen, private communication
\end{thebibliography}
\end{document}